\documentclass[12pt]{article}
    \usepackage{amsmath,amsxtra,amssymb, amsfonts, bbm} 

 \def\be{\begin{eqnarray}}
\def\ee{\end{eqnarray}}
\def\bee{\begin{eqnarray*}}
\def\eee{\end{eqnarray*}}

\def\pmx{\begin{pmatrix}}
\def\emx{\end{pmatrix}}
\def\bsq{\begin{subequations}}
\def\esq{\end{subequations}}

\newtheorem{thm}{Theorem}
\newtheorem{cor}[thm]{Corollary}
\newtheorem{lemma}[thm]{Lemma}

\newtheorem{prop}[thm]{Proposition}

 \newcommand{\proj}[1]{ | #1 \kb  #1|}
 
 \def\ts{\textstyle}

\def\ot{\otimes}
\def\tr{{\rm Tr} \, }
\def\trp{{\rm Tr} }
\def\bra{\langle}
\def\ket{\rangle}
\def\kb{ \ket \bra }
\def\raw{\rightarrow}
\def\imp{\Rightarrow}
\def\iff{\Leftrightarrow}
\def\half{\tfrac{1}{2}}
\def\wtd{\widetilde}
\def\nn{\nonumber}

\def\pf{\noindent {\bf Proof:} ~ }
\def\qed{\quad {\bf QED}}
\def\wh{\widehat}

\def\ca{{\cal A}}
\def\ck{{\cal K}}

\def\cw{{\cal W}}

\def\ch{{\cal H}}
\def\cb{{\cal B}}
  \def\one{{\mathbbm 1}}

\title{\Large A Unified Treatment of  Convexity of Relative Entropy and Related Trace Functions, with Conditions for Equality}

\author {Anna Jen\v{c}ov\'{a} \thanks{Supported by the grants VEGA 2/0032/09,
APVV- 0071-06, Center of Excellence SAS - Quantum Technologies
and ERDF OP R\&D Project CE QUTE ITMS  26240120009}\\
{\small Mathematical Institute, Slovak Academy of Sciences,}\\
{\small \v {S}tef\'{a}nikova 49, 814 73 Bratislava, Slovakia} \\
{\small  jenca@mat.savba.sk}  \\  ~~  \\ \and  
Mary Beth Ruskai \thanks{Partially supported by  National Science Foundation 
 under Grant DMS-0604900.}
 \\{\small  Department of Mathematics, Tufts University, Medford,
MA 02155, USA} \\ {\small  marybeth.ruskai@tufts.edu} }

\begin{document}

\maketitle

\begin{abstract}
We consider a generalization of relative entropy derived from the 
Wigner-Yanase-Dyson entropy and give a
simple, self-contained proof that it is convex.  Moreover, special cases yield
the joint convexity of relative entropy, and for $\tr K^* A^p K B^{1-p} $ Lieb's
joint concavity in $(A,B) $ for   $0 < p < 1$ and Ando's joint convexity for $1 < p \leq 2$.
This approach allows us to obtain conditions for equality in these cases, as
well as conditions for equality in a number of inequalities which follow from them.
These include the monotonicity under partial traces, and some Minkowski type
matrix inequalities proved by Carlen and Lieb for $ \trp_1 ( \trp_2\,  A_{12}^p)^{1/p}$.   
In all cases  the equality conditions are independent of $p$;  for extensions to 
three spaces they are identical to the conditions for equality in the
strong subadditivity of relative entropy. 
\end{abstract}

\pagebreak

\section{Introduction}

\subsection{Background}

For matrices  $A_{12} > 0$ acting on a tensor product of two Hilbert spaces,
 Carlen and Lieb  \cite{CL1,CL2} considered the trace function 
 $  \big[ \trp_1 ( \trp_2 A_{12}^p)^{q/p} \big]^{1/q}$  and proved
that it is concave when  $0 \leq p \leq q \leq 1$ and  convex when
$1 \leq q$ and $1 \leq p \leq 2 $.  They showed that this implies that these functions and the norms
they generate satisfy Minkowski type inequalities, including a    natural
generalization to matrices $A_{123}$ acting on a tensor product of three
Hilbert spaces.    They also raised the question  
of the conditions for equality in their inequalities.   
When $ q = 1$, we show that this can be treated using methods
developed to treat equality in the strong subadditivity of quantum entropy.
Moreover, we obtain conditions for equality in a large class of related convexity
inequalities, show that they are independent of $p$ in the range $0 < p < 2$,
and show that for inequalities involving $A_{123}$ they are identical to the equality 
conditions for strong subadditivity (SSA) of quantum entropy given in \cite{HJPW}  

These equality conditions are non-trivial and have
found many applications in quantum information theory.   For
example, they play an  important role in some recent   ``no broadcasting''
results; see \cite{LLC} and references therein.   They also play a key role 
in Devetak and Yard's \cite{DY} ``quantum state redistribution'' protocol
which gives an  operational interpretation to the 
quantum conditional mutual information.

Our approach to proving joint convexity of relative entropy is motivated by 
 Araki's relative modular operator \cite{Ak}, introduced to generalize 
 relative entropy to more general situations including type III von Neumann algebras.
It was subsequently used by  Narnhofer and Thirring  \cite{NT}  to give a new 
proof of SSA.
The argument given here is similar to that in \cite{LesR,Pz1,Rssa};  
however, the unified treatment for $0 < p < 2$
leading to equality conditions, is new.   Moreover, a dual treatment can be given for
$-1 < p < 1$ allowing extension to the full range $(-1,2)$.   

Wigner and Yanase  \cite{WY,WY2}  introduced the notion of skew information of
a density matrix $\gamma$ with respect to a self-adjoint observable $K$,
\be  \label{skew}
   -  \tr \half [K,\gamma^p] \, [K, \gamma^{1-p}]
\ee
for $p = \half$ and Dyson suggested extending this to $p \in (0,1)$.
Wigner and Yanase \cite{WY2} proved that \eqref{skew} is convex in $\gamma$
for $p = \half$ and, in his seminal paper \cite{Lb} on convex
trace functions, Lieb proved joint concavity  for $p \in (0,1)$ for the more
general function
\be    \label{wyd}
    (A,B) \mapsto \tr  K^* A^p K B^{1-p}
\ee
for $K$ fixed and $A,B > 0$ positive semi-definite.  This implies convexity
of \eqref{skew} and
 was a key step in the original proof \cite{SSA}
of the strong subadditivity (SSA) inequality of quantum entropy.   Moreover, it
leads to a proof of joint convexity of relative entropy\footnote{ In 
\cite{SSA} only concavity of the conditional entropy was proved explicitly,
but the same argument \cite[Section~V.B]{MBR} yields joint convexity of the relative entropy.
Independently, Lindblad \cite{Lind74} observed that this follows
directly from \eqref{wyd} by differentiating at $p = 1$.} as well.
It is less well known that  Ando \cite{Ando1,Ando2} gave another proof
which also  showed that for $1 \leq p \leq 2$,
the function \eqref{wyd}  is jointly convex in $A,B$.   The case $p = 2$
was considered earlier by Lieb and Ruskai \cite{LbR2}.
We modify what one might describe as Lieb's extension of the Wigner-Yanase-Dyson (WYD)
entropy to a type of relative entropy  in a way that allows a unified treatment of the convexity and
concavity of  $\tr  K^* A^p K B^{1-p} $ in the range  $p \in (0,2]$ and
  includes the usual relative entropy as a special case.    Our modification retains
 a linear term, even for $A \neq B$.   Although this might seem unnecessary 
 for  convexity and concavity questions, it is crucial to a unified treatment.

Lieb also considered $\tr  K^* A^p K B^{q} $ with $p, q > 0$ and $0 \leq p + q \leq  1$
and Ando considered   $1 < q \leq p \leq 2$.    In Section~\ref{sect:qneq1}, we extend
our results to this situation.   However, we also show that for $q \neq 1-p$, equality
holds only under trivial conditions.  Therefore, we concentrate on the case $q = 1-p$.

Next, we introduce our notation and conventions.   In Section~\ref{sect:WYDgen},
we first describe our generalization of relative entropy and prove its convexity;
then consider the extension to $q \neq 1-p$ mentioned above; and finally prove
 monotonicity under partial traces including a generalization of strong subadditivity
 to $p \neq 1$.    In Section~\ref{sect:eq}, we consider several formulations of
 equality conditions.  In Section~\ref{sect:LC}, we show how to use these results
 to obtain equality conditions in the results of Lieb and Carlen \cite{CL1,CL2}.
For completeness, we include an appendix which contains the proof of
 a basic convexity result from \cite{Rssa} that is key to our results.

\subsection{Notation and conventions}  \label{sect:not}

We introduce two linear maps on the space $M_d$ of $d \times d$ matrices.  
Left multiplication by $A$ is denoted $L_A$ and defined as $L_A(X) = AX$;
right multiplication by $B$ is denoted $R_B$ and defined as $R_B(X) = XR$.
These maps are associated with the relative modular operator 
$\Delta_{AB} = L_A R_B^{-1}$  introduced by Araki in a far more general context.
They have the following properties:

\begin{enumerate}

 \renewcommand{\labelenumi}{\theenumi}
    \renewcommand{\theenumi}{\alph{enumi})}
    
    \item The operators $L_A$ and $R_B$ commute since
    \be
      L_A [R_B(X) ] =  AXB = R_B[L_A(X)]
    \ee
even when $A$ and $B$ do not commute.

\item  $L_A$ and $R_A$ are invertible if and only if $A$ is non-singular,
in which case 
 $L_A^{-1} = L_{A^{-1}}$ and $R_A^{-1} = R_{A^{-1}}$.
 
 \item  When $A$ is self-adjoint, $L_A$ and $R_A$ are both self-adjoint with
 respect to the Hilbert Schmidt inner product $\bra A, B \ket = \tr A^* B$.
 \item  When $A \geq 0$, the operators $L_A$ and $R_A$ are positive 
 semi-definite, i.e., 
  \bee
    \tr X^{*} L_A(X) =   \tr X^{*} AX \geq 0 \qquad \text{and} \\
          \tr X^{*} R_A(X) =   \tr X^{*} X A = \tr  X A X^{*} \geq 0.
 \eee 
 
 \item When $A  > 0$, then $(L_A)^p = L_{A^p}$ and  $(R_A)^p = R_{A^p}$ for all $p \geq 0$.
   If $A$ is also non-singular,  this extends to all $p \in {\bf R}$.   More generally
   $f(L_A) = L_{f(A)}$ for $f: (0,\infty ) \mapsto {\bf R}$.

\end{enumerate}

To see why (e) holds, it suffices to observe that $A > 0$ implies
$L_A$ and $R_A$ are linear operators for which $f(A)$ can be defined by
the spectral theorem for any function $f$ with domain in $(0, \infty)$.
It is easy to verify that $A | \phi_j \ket = \alpha_j | \phi_j \ket $ implies
$L_A | \phi_j \kb \phi_k | = \alpha_j | \phi_j \kb \phi_k |$ for $ k = 1 \ldots d$
so that the  spectral decomposition of $A$ induces one on $L_A$ with
degeneracy $d$ and $f(L_A)  | \phi_j \kb \phi_k |  = f(\alpha_k)  | \phi_j \kb \phi_k | $.
For $R_B$ a similar argument goes through starting with left eigenvectors of $B$
i.e.,  $ \bra \phi_j | B = \beta_j  \bra \phi_j | $.

If a function is homogeneous of degree 1, then convexity is equivalent
to subadditivity.   Thus, if 
$F(\lambda A) = \lambda F(A)$, then $F$ is convex
if and only if $F(A) \leq  \sum_j F(A_j)$ with $A = \sum_j A_j$.
 We will use this equivalence without further ado.
 
For $B$ positive semi-definite, we denote the projection onto $(\ker B)^\perp$ by 
 $P_{ (\ker B)^\perp}$.
We will encounter expressions involving commuting positive semi-definite matrices
$A, D$ with $\ker D \subseteq \ker A$.   We will simply write $A D^{-1}$ for
\be   \lim_{\epsilon \raw 0} \sqrt{A} (D + \epsilon I)^{-1} \sqrt{A} = A D^{-1} P_{ (\ker D)^\perp} =  A D^{-1} P_{ (\ker A)^\perp} 
\ee 
with $D^{-1}$ the generalized inverse.

\section{WYD entropy revisited and extended}  \label{sect:WYDgen}

\subsection{Generalization of relative entropy}

 We now introduce the family of functions
 \be  \label{gp}
       g_p(x) =  \begin{cases} \frac{1}{p(1-p)} (x - x^p)  & p  \neq 1
                                       \\   x \log x  &   p = 1 \end{cases}.
 \ee
 which are well-defined for $x > 0$ and $p \neq 0$.   We will consider $p \in (0,2]$
although it would suffice to consider  $p \in [\half,2]$.
 For $A, B$ strictly positive we define 
  \be   \label{quasi}       J_p(K,A,B)  & \equiv & \tr \sqrt{B} K^*  \, g_p\big(L_A R_B^{-1}\big)  (K  \sqrt{B})  \\
         & =  & \begin{cases}
           \frac{1}{p(1-p)}   \big(  \tr K^* A K - \tr  K^* A^{p} K B^{1-p}  \big)&  p \in (0,1) \cup   (1,2)
              \\ \tr   KK^* A  \log A -  \tr K^* A K  \log B   & p = 1  \\
            -\half  \big(  \tr K^* A K - \tr A K B^{-1} K^* A\big)    & p = 2 \end{cases} \label{Jpdef}
  \ee
When $p = 1$ and $K = I$, \eqref{quasi} reduces to the usual
relative entropy, i.e.,
\be
   J_1(I,A,B) = H(A,B) = \tr A ( \log A - \log B )
\ee
For $ p \neq 1$, the function $J_p(K,A,B) $ differs from that considered by
Lieb \cite{Lb} and Ando  \cite{Ando1,Ando2} by the seemingly irrelevant
linear term $ \tr K^* A K $ and the factor $   \frac{1}{p(1-p)} $.   However, this
minor difference allows us to give a unified treatment of $p \in (0,2]$ because
of the  extension by continuity to $p = 1$ and the sign change there.

One might expect to associate the exchange $A \leftrightarrow B$ with
the symmetry $p  \leftrightarrow (1-p)$ around $p = \half$.   However,
there are several subtleties due to the linear term, the exchange $K \leftrightarrow K^*$,
and the case $p = 1$.   Therefore, we use instead the observation
that
\be   \label{Jt}
   J_p(K^*, B,A)  & = &   \tr \sqrt{A}  \, K \, g_p\big(L_B R_A^{-1}\big) (K^* \sqrt{A})   \nn  \\
      & = & \tr \sqrt{B} K^* \,  \wtd{g}_{1-p}\big(L_A R_B^{-1}\big) (K \sqrt{B})  \nn  \\
      & = & \wtd{J}_{1-p}(K,A,B)
\ee
 where, for   $-1 \leq p < 1$, we define
  \be \label{wtdg}
     \wtd{g}_{p}(x) = x g_{1-p}(x^{-1}) =
       \begin{cases} \frac{1}{p(1-p)} (1 - x^p ) & p \neq 0
                                \\   - \log x  &   p = 0 \end{cases}
\ee
and $\wtd J_p(K,A,B) = 
 \tr \sqrt{B} K^*  \, \wtd g_p\big(L_A R_B^{-1}\big)  (K  \sqrt{B}) $.
 
  The functions $J_p(K,A,B)$ and $\wtd{J}_p(K,A,B)$ 
  have been considered before, usually with $K = I$,
in the context of information geometry (\cite[Section 7.2]{AN} and references
therein)  and by Petz \cite{Pz1} who used the term ``quasi-entropy''.   
What is novel here is that we
present a simple unified proof of joint convexity in $A,B$ that easily yields
equality conditions, shows that they are independent of $p$, and can be
extended to other functions.

The special case $J_p(I,A,I)$ is equivalent\footnote{This was pointed out by Karol Zyczkowski}
 to the Tsallis \cite{T} entropy.
 When $K = K^*$, the relation
 \be
     J_p(K,A,A) = -  \frac{1}{2p(1-p)} \tr [K,A^p][K,A^{1-p}]
 \ee
  yields the original WYD information (up to a constant) and extends it to the range $(0,2]$.
  Morevoer,   $K = K^*$ implies that   $ J_p(K,A,A) = \wtd{J}_{1-p}(K,A,A)$.
     Although neither $g_p(w)$ nor $\wtd{g}_{1-p}(w)$ is positive, their 
      average\footnote{The definition of $\wtd{g}_p$ in \eqref{wtdg} differs from
     that in \cite{LesR}  by the exchange   $ \wtd{g}_p  \leftrightarrow  \wtd{g}_{1-p}  $ so that
   in \cite{LesR}  $G(w) = \half [g(w) + \wtd{g}(w)]$ for any $g$.  In the
     convention used here,  $G_p(w) = \half [g_p(w) + \wtd{g}_{1-p}(w)]$.} 
   
        $G_p(w) \equiv \half [g_p(w) + w g_p(w^{-1})] \geq 0 $ on $(0,\infty)$.
        Therefore,  when $K = K^*$,
  \be
       J_p(K,A,A) =  \tr (K \sqrt{A} )^* G_p(L_A R_A^{-1} )(K \sqrt{A})  \geq 0
  \ee

  The function $J_p(I,A,B) $ is   a more appealing
  generalization of relative entropy than $\tr A^p B^{1-p}$ because of  Proposition~\ref{pseudo},
  which one can consider to be a generalization of Klein's inequality \cite{Klein}.
  It allows one to use $J_p(I,A,B) $ as a pseudo-metric, as is commonly
  done with the relative entropy.    

\begin{prop}  \label{pseudo}
 When $U$ is unitary and $A,B > 0$ with $\tr A = \tr B = 1$, then $J_p(U,A,B)   \geq 0$  with
 equality if and only if $A= U^* B U$.
\end{prop}
\pf  When $U$ is unitary,
\be   \label{unit}
    J_p(U,A,B)  = J_p(I,U^*AU,B) = J_p(I,A,UBU^*) .
\ee
Therefore, it suffices to consider the case $U = I$.   
For $p \in (0,1)$ H\"olders inequality implies
$\tr A^p B^{1-p} \leq (\tr A)^p (\tr B)^{1-p} = 1$ with equality if and
and only $A =B$.   It immediately follows that
\be   \label{klein}
   J_p(I,A,B)  \geq    \tfrac{1}{p(1-p)} \big(\tr A - 1 \big) = 0 \qquad   \hbox{and}  \qquad
    J_p(I,A,B )= 0  \iff A = B .
\ee
For $p = 1$, the result is well-known \cite[Section 2.5.2]{Ruelle} and 
  originally due to O. Klein \cite{Klein}.
For $p \in (1,2)$ we write $p = 1+r$ and again use H\"older's inequality
\be
     1 & =  & \tr A   =  \tr B^{-\frac{r}{2(r+1)} }  A B^{-\frac{r}{2(r+1)} } B^{\frac{r}{r+1} }   \nn  \\
         & \leq &   \Big[  \tr \big( B^{-\frac{r}{2(r+1)}}   A B^{-\frac{r}{2(r+1)} }  \big)^{1+r} \Big]^{\frac{1}{1+r} }   \, (\tr B)^{\frac{r}{1+r} }\\  \nn
         & \leq & \big[ \tr B^{-\half} A^{1+r} B^{-\half} \big]^{\frac{1}{1+r} } \big( \tr A^{1+r} B^{-r} \big)^{\frac{1}{1+r} }
\ee
 where we used $\tr B = 1$ and the second inequality follows from a classic result
 of Lieb-Thirring \cite[Appendix B, Theorem 9]{LT}.   \qed

Because the denominator $p(1-p)$ changes sign at $p = 0$ and $p =1$,
both $g_p$ and $\wtd{g}_p$ are convex.    In fact, they satisfy
the much stronger condition of  operator convexity for  $p \in (0,2]$ and
$p \in [-1,1)$ respectively.
Since $g(0) = 0$ and
\be
     \frac{ g_p(x) }{x} = \begin{cases}  \frac{1}{p(1-p)} (1 - x^{p-1}) & p  \neq 1
                                \\     \log x  &  p = 1\end{cases},
\ee
it follows  that  $ g_p(x) / x $ is operator monotone \cite{Ando1,Don,Lw},
for $p \in (0,2]$, i.e.,  
 $g_p$ can be analytically continued to the upper half plane, which
it maps into itself.   By applying Nevanlinna's theorem 
\cite[Section 59, Theorem 2]{AG}  to
$g_p(x)/x$, one finds
that $g_p(x)$ has an integral representation of the form
\be  \label{intrep}
      g_p(x) & = &  ax + \int_0^{\infty}     \frac{x^2t - x}{x + t}  \,  d\nu(t)   \nn \\
          & = & ax +
          \int_0^{\infty} \Big[ \frac{x^2  }{x+t} - \frac{1}{t}  + \frac{1}{x+t}  \Big]  \, t  \, d\nu(t)  
\ee
 with $\nu(t) \geq 0 $.   Integral representations are not unique, and making a suitable change of variable in the classic formula
 \be
    \int_0^\infty   \frac{x^{p-1} }{x+1} = \frac{\pi}{\sin p \pi}  \equiv \frac{1}{c_p}  \qquad   \qquad p \in (0,1)
 \ee
allows us to give  the following explicit representations
\be   \label{intspec}
      g_p(x) = \begin{cases}    \frac{1}{p(1-p)} \Big[ x + c_p  \int_0^{\infty}
                   \big( \frac{t}{x+t} - 1 \big)  t^{p-1} dt  \Big]  & p \in (0,1) \\  ~ & ~ \\
                   \int_0^{\infty} \big( \frac{x^2}{x+t} - 1 + \frac{t}{x+t} \big) \frac{1}{1+t} dt& p = 1    \\ ~ & ~ \\
                   \frac{1}{p(1-p)} \Big[ x -   c_{p-1} \int_0^{\infty}  \frac{x^2}{x+t}  t^{p-2} dt   \Big] & 
                   p \in (1,2) \\ ~ & ~ \\
                   \half (- x  + x^2) & p = 2   \end{cases}
\ee
Note that for $p \in (0,2)$
the integrand is supported on $(0, \infty)$.  This plays a key
role in the equality conditions; therefore, we will henceforth concentrate on $p \in (0,2)$.

 \begin{thm}   \label {thm:Jp}
The function $J_p(K,A,B) $ defined in \eqref{quasi} is jointly convex in $A,B$.
 \end{thm}
\pf  It follows from \eqref{intrep} that
\begin{align}   \label{intAB}
 J_{p}(K,A,B)   & ~  =  ~ a \tr K^* A K       \\
          +  \int_0^{\infty}   & \Big[ \tr  K^*  A  \frac{1}{L_A + t R_B} (A K)   -  \frac{\tr K B K^*}{t}  +
        \tr B K^* \frac{1}{L_A + t R_B} (K B)  \Big]   t  \, \nu(t) \, dt   \qquad \nn
\end{align}
The joint convexity then follows immediately from that of the map
$(X,A,B) \mapsto   \tr X^* \frac{1}{L_A + t R_B} (X) $ which was proved
in \cite{Rssa} following the strategy in \cite{LbR2}.  
The proof is also given in the Appendix.  \qed

For other approaches see Petz \cite{OP,Pz1}, Effros \cite{Ef},  The advantage to the argument used
here is that it immediately implies that equality holds in  joint convexity
if and only if it holds for each term in the integrand.
    
  \begin{cor}
  The relative entropy  $H(A,B) = J_1(I,A,B) $  is jointly convex in $A,B$.
  \end{cor}

\subsection{Extensions with $ r \neq 1-p$.}  \label{sect:qneq1}

 We now consider extensions of Theorem~\ref{thm:Jp} to situations 
 considered by Ando \cite{Ando2} and Lieb \cite{Lb}  in which
 $B^{1-p}$ is replaced by $B^r$ with $r \neq 1-p$.   Our approach uses an idea
 from Bekjan \cite{Bek} and  Effros \cite{Ef}.
 We will also show that equality holds in these extensions only under trivial conditions.
 For this we first need an elementary lemma, which we prove for the concave case.
 \begin{lemma}  \label{lemma:strict}
 Let  $f(\lambda):[0,\infty) \mapsto {\bf R}$ be  a non-linear convex or concave operator
function, let $A_1, A_2$ be
density matrices and $A =  \lambda A_1 + (1-\lambda) A_2$ with $\lambda \in (0,1)$.  Then
 $f(A) = \lambda f(A_1) + (1-\lambda) f(A_2)$
 if and  only if $A_1 = A_2$.
 \end{lemma}
 \pf  Since any operator concave function is analytic, non-linearity
 implies that $f$ is strictly concave.    If $f(A) = \lambda f(A_1) + (1-\lambda) f(A_2)$,
 then
 \be
    \bra v, f(A)  v \ket = \lambda  \bra v, f(A_1)  v \ket  + (1-\lambda)  \bra v, f(A_2)  v \ket
 \ee
 for any vector $v$.   Now choose $v$ to be a normalized eigenvector of $A$.  Then
 inserting this on the left above and applying Jensen's inequality to each
 term on the right, one finds
 \be
   f\big(  \bra v, A v \ket \big)  \leq \lambda  f\big(  \bra v, A_1 v \ket \big) + (1-\lambda)  f\big(  \bra v, A_2 v \ket \big)
 \ee
 But this contradicts concavity unless equality holds, which implies that
 $v$ is also an eigenvector of $A_1$ and $A_2$.   But then the strict
 concavity of $f$ also implies that   $\bra v, A_1 v \ket =  \bra v, A_2 v \ket $.
 Since this holds for an orthonormal basis of eigenvectors of $A, A_1$ and $A_2$, we
 must have $A_1 = A_2$.

 \begin{cor}  \label{cor:p<1}
The function $(A,B)\mapsto \tr K^*A^p KB^r$ is jointly concave on the
set of positive definite matrices when $p,r \geq 0$ and $p+r \leq 1$.
  Moreover, when $p + r < 1$ and $K$ is invertible, the convexity is strict 
  unless $B_1 = B_2$ and $A_1 =A_2 $.
 \end{cor}
 \pf  It is an immediate consequence of Theorem~\ref {thm:Jp} that
 $(A,B) \mapsto \tr K^*A^p K B^{1-p}$ is jointly concave in $A,B$.   Now write
$   \tr K^*A^p KB^r = \tr K^*A^p K(B^s)^{1-p}$ with $s = r/(1-p)$.
First, observe that for $0 < s  < 1$ the function $f(x) = x^s$ satisfies
the hypotheses of Lemma~\ref{lemma:strict}.   Therefore,
\be  \label{strict1}
   \big(\lambda B_1 + (1-\lambda) B_2\big)^s > \lambda B_1^s + (1-\lambda) B_2^s
\ee
with $0 < \lambda < 1$ and $B_1 \neq B_2 $.
The  operator monotonicity of  $x\mapsto x^{1-p}$ for $0 < p < 1$ then
implies
\be  \label{strict2}
   \big(\lambda B_1 + (1-\lambda) B_2\big)^r  > \big(  \lambda B_1^s + (1-\lambda) B_2^s  \big)^{1-p},
\ee
and the joint concavity  of  $\tr K^*A^p K B^{1-p}$ implies
\be \label{concpr}
    \tr K^*A^p K(B^s)^{1-p} &  \geq &  \tr K^*\big(\lambda A_1 + (1-\lambda) A_2\big)^p
   K \big(\lambda B_1^s + (1-\lambda) B_2^s \big)^{1-p}   \\
         & \geq & \lambda  \tr K^*A_1^p KB_1^{s(1-p)} + (1-\lambda) \tr K^*A_2^p KB_2^{s(1-p)}     \nn
\ee
where $A=\lambda A_1+(1-\lambda)A_2$, $B=\lambda B_1 +(1-\lambda)B_2$,
which is precisely the joint concavity of  $\tr K^*A^p KB^r$.
Moreover,   equality in joint concavity implies equality  in \eqref{concpr}
 and, since $K^*A^pK$ is strictly positive, this implies equality in
\eqref{strict1}.     Therefore,
equality in \eqref{concpr}  gives a contradiction unless $B_1 = B_2$.  
In that case, the joint concavity reduces to concavity
  in $A$ for which, by a similar argument, equality holds if and only if
  $A_1 =A_2 $. \qed

   \begin{cor}  \label{cor:p>1}
The function $(A,B)\mapsto \tr K^*A^p KB^{1-r}$ is jointly convex on the
set of positive definite matrices when $1 < r \leq p \leq 2$.  Moreover, 
when $r < p$ and $K$ is invertible, the convexity is strict unless
  $B_1 = B_2$ and $A_1 =A_2 $.
 \end{cor}
 \pf   The argument is similar to that for Corollary~\ref{cor:p<1}.
Write $   \tr K^*A^p KB^{1-r} = \tr K^* A^p K(B^{s})^{1-p}$ with $s = \frac{1-r}{1-p}$.
Since   $s \in (0,1)$ and $ 1-p \in (-1,0) $ when  $1 < r  < p < 2$, it follows that
$x^{s}$ is operator concave and $x^{1-p}$ is operator monotone decreasing.
\qed

 \subsection{Monotonicity under partial traces}  \label{sect:mono}

Let $X $ and $Z $ denote the generalized Pauli operators whose action
on the standard basis is  $X  |e_k\ket =  |e_{k+1}\ket$ (with subscript addition mod $d$)
and $Z |e_k\ket   = e^{i2 \pi k  /d}  |e_k\ket $.
 It is well known and easy to verify that  $\tfrac{1}{d} \sum_k Z^k A Z^{-k}$ is the
 projection of a matrix onto its diagonal.   If $D$ is a diagonal matrix,
 then  $ \sum_k X^k D X^{-k} = (\tr D) I$.   Now let $\{ W_n \}_{n = 1,2 \ldots d^2}$
 denote some ordering of the generalized Pauli operators , e.g.,
 $W_{j + k(d-1) } =   X^j Z^k$ with $j, k = 1,2 \ldots d$.  Then
 $       \tfrac{1}{d} \sum_n W_n A  W_n^* = (\tr A) I $ and
\be    \label{vw} 
    \tfrac{1}{d} \sum_n (W_n\ot I_2)  \, A_{12} \,  (W_n\ot I_2)^*  = I_1\ot
    (\trp_1 \, A) = I_1 \ot A_2
 \ee
 Using the fact that replacing $W_n$ by $U W_n U^*$ with $U$ unitary,
 simply corresponds to a
 change of basis which does not affect \eqref{vw} and then multiplying both
 sides by $U^* \ot I_2 $ on the left and $U \ot I_2$ on the right gives the
 equivalent expression
 \be    \label{vw2} 
    \tfrac{1}{d} \sum_n (W_n U^* \ot I_2)  \, A_{12} \,  (W_n U^* \ot I_2)^*  =  
     I_1 \ot A_2
 \ee
 Combining this with joint convexity yields a slight generalization of the well-known 
 monotonicity of 
$J_p(K,A,B)$  under partial traces (MPT), first proved by Lieb in \cite{Lb} for the
 case $K_{12} = I_1 \ot K_2$ when $p \in (0,1)$.

\begin{thm}  \label{thmJpmono}
Let $J_p$ be as in \eqref{Jpdef},
$ A_{12}, B_{12}$ strictly positive in $M_{d_1} \ot M_{d_2}$
and $K_{12}  = V_1 \ot K_2$ with $V_1$ unitary in $M_{d_1}$.
Then
\be  \label{Jpmono}
    J_p(K_2,A_2,B_2) \leq  J_p(K_{12}, A_{12}, B_{12} )
\ee
 \end{thm}  
 \pf   Writing $\cw_n $ for $W_n \ot I_2$ and using \eqref{vw2} gives 
     \bee 
       J_p(K_2,A_2,B_2)  & = &     \tfrac{1}{d_1}  J_p(I_1\ot K_2 ,I_1  \ot A_2, I_1  \ot B_2) \label{ft0}    \\ \nn
        & = &   \tfrac{1}{d_1}   \textstyle{ J_p \Big( I_1 \ot K_2 , \,
       \tfrac{1}{d_1 } \sum_n \cw_n  (V_1^* \ot I_2) A_{12}   (V_1 \ot I_2)  \cw_n^* ,~
         \tfrac{1}{d_1 }  \sum_n   \cw_n B_{12}    \cw_n^* \Big) } \\  \label{fta}
              & \leq &  \tfrac{1}{d_1^2} \sum_n J_p\big(I_1 \ot K_2 ,
       \cw_n  (V_1^* \ot I_2) A_{12}   (V_1 \ot I_2)  \cw_n^* ,   \cw_nB_{12}   \cw_n^* \big) \\ 
            & = &   J_p(V_1 \ot K_2 , A_{12}, B_{12} )     
 \eee
 where  the final equality follows
 from the unitary invariance of the trace.   \qed

Because $\tr_{12} (V_1 \ot K_2) A_{12}  (V_1 \ot K_2)^* = \tr_2 K_2 A_2  K_2^* $, \eqref{Jpmono}
is equivalent to  
\be
    \tr K_2^* A_2^p K_2 B_2^{1-p} - \tr  (V_1 \ot K_2)^* A_{12}^p  (V_1 \ot K_2) B_{12}^{1-p} 
    \begin{cases}  \geq 0 & p \in (0,1) \\ \leq 0 & p \in (1,2) \end{cases}.
\ee
 We can obtain a weak reversal of this for $p \in (0,1)$.
 The argument in the Appendix shows that for any $p$ and fixed $A,B \geq 0$ both
 $\tr K^* A^p K B^{1-p} $ and $\tr K^* A K$ are convex in $K$.  This was observed earlier by Lieb \cite{Lb}
 and also follows from the results in \cite{LbR2}.   One can then apply the argument above
 in the special case $A_{12} = I_1 \ot A_2 ,  B_{12} = I_1 \ot B_2$ to conclude that
\be     \label{rev}
 \tr K_{2}^* A_2^p K_2 B_2^{1-p} &  \leq  &  \tfrac{1}{d_1} \tr K_{12}^* (I_1 \ot A_2)^p K_{12}  (I_1 \ot B_2)^{1-p}  \\
             & \leq &  \tr K_{12}^* (I_1 \ot A_2)^p K_{12}  (I_1 \ot B_2)^{1-p} 
\ee
independent of whether $p < 1$ or $p > 1$.
However, because the term $\tr K^* A K $
 is  convex rather than linear in $K$, \eqref{rev} does not
allow us to draw any conclusions about the monotonicity of  
$J_p(K_{12}, I_1 \ot A_2, I_1 \ot B_2)$.

To prove Theorem~\ref{thmJpmono}
 we showed that joint convexity implies monotonicity;
the reverse implication also holds.  
Let $A_1,\dots,A_m$, $B_1,\dots,B_m$ be positive definite
matrices in $M_d$, $A=\sum_jA_j$, $B=\sum_jB_j$,
and put
\be\label{ABdiag}
\wtd{A}_{12}=\sum_j \proj{e_j}\ot A_j, \quad \wtd{B}_{12}=\sum_j \proj{e_j}\ot B_j,
\ee
for $e_1,\dots,e_m$ the standard basis of ${\bf C}^m$.
Then $\wtd{A}_{12}$ and $\wtd{B}_{12}$  are block diagonal, and
$\wtd{A}_2 = \trp_1 \wtd{A}_{12} = \sum_k A_k = A$  and similarly for $B$.
Then if monotonicity under partial traces holds, one can conclude that
\be\label{conv_monot}
J_p( K,A,B) &=& J_p(K,\wtd{A}_2,\wtd{B}_2)\notag\\
&\le&  J_p(I_1\ot K,\wtd A_{12},\wtd B_{12})=\sum_j J_p(K,A_j,B_j)
\ee
Thus, monotonicity under partial traces also directly
implies  joint convexity of $J_p$.    

   Applying \eqref{Jpmono} in the case $K = I$,  and $A_{12} \mapsto A_{123}$ and
    $B_{12} \mapsto  A_{12} \ot I_3 $ gives
      \be    \label{Jpssa}
       J_p(I_{23} , A_{23},   A_2 \ot I_3) \leq J_p(I_{123} , A_{123},  A_{12} \ot I_3)
    \ee
When $p = 1$, it follows from \eqref{Jpdef}  that
\bee
    J_1(I_{23} , A_{23},    A_2 \ot I_3) =  H(A_{23},  A_2 \ot I_2) = -S(A_{23}) + S(A_2)
\eee
where  $S(A) = - \tr A \log A$.  Thus, \eqref{Jpssa} becomes
\bee
    -S(A_{23}) + S(A_2)   \leq  -S(A_{123}) + S(A_{12})
\eee
or, equivalently
\be  \label{ssa}
    S(A_2)  +  S(A_{123})  \leq  S(A_{12}) + S(A_{23})
\ee
which is the standard form of SSA.


\section{Equality  for joint convexity of  $J_p(K,A,B)$.}  \label{sect:eq}

\subsection{Origin of necessary and sufficient conditions}    \label{sect:Jpeqal}
Looking back at the proof of Theorem \ref{thm:Jp}, we see that
for $p \in (0,2)$, equality holds in the joint convexity of $J_p(K,A,B)$
if and only if equality holds in the joint convexity   for each term in
the integrand in \eqref{intrep}.  It should be clear from the argument given
in the Appendix, that this requires $M_j = 0$ for all $j$ with $M_j$
given by \eqref{Mj}.   This is easily seen to be equivalent to
\be   \label{basiceq}
   (L_{A_j} + t R_{B_j})^{-1}(X_j)  =   (L_A + t R_B)^{-1}(X)  \quad \text{for all} ~ j.
\ee
with $A = \sum_j A_j, B = \sum_j B_j $, and $X = \sum_j X_j  $ with $X_j = A_j K$ and/or $X_j = K B_j$.
By writing $A K = L_A(K)$ in the former case and $KB = R_B(K)$ in the latter
we obtain the conditions
 \bsq
 \be
     ( I + t  \Delta_{A_j B_j }^{-1} )^{-1}(K) =    ( I + t  \Delta_{AB}^{-1} )^{-1}(K)
        \quad  \forall ~j \quad \forall t > 0     \label{delta} \\
     (  \Delta_{A_j B_j } + t I)^{-1}(K) =   (  \Delta_{A B  } + t I)^{-1}(K)
     \quad  \forall ~j \quad \forall t > 0   \label{deltb}
     \ee     \esq 
     From the integral representations \eqref{intspec}, one might expect
     it to be necessary for either or both of \eqref{delta}  and \eqref{deltb}  to hold
     depending on $p$.
     In fact, either will suffice because \eqref{delta} holds if and only if \eqref{deltb} holds.
Because $\Delta_{AB}$ is
positive definite,  by analytic continuation \eqref{deltb} extends from $t> 0$
to the entire complex plane, except points $-t$ on the negative real axis for
which $t \in $ spectrum$(\Delta_{AB})$.   Therefore, by using the Cauchy integral formula,
one finds that for any function $G$ analytic on ${\bf C} \backslash (- \infty,0]$
$G(\Delta_{A_j B_j })(K) = G(\Delta_{AB})(K)$.


 \begin{thm}   \label{thm:eqJp}
 For fixed $K$, and $A = \sum_j A_j, B = \sum_j B_j$,  the following are equivalent

 a) $ J_p(K,A,B) =\sum_j  J_p(K,A_j,B_j)$  for all $p \in (0,2)$.

 b) $ J_p(K,A,B) =\sum_j  J_p(K,A_j,B_j)$  for some  $p \in (0,2)$.

 c) $ (  \Delta_{A_j B_j } + t I)^{-1}(K) =   (  \Delta_{A B  } + t I)^{-1}(K) $ for all $j$ and for all $t > 0$.

 d)  $A_j^{it} K B_j^{-it} = A^{it} K B^{-it} $  for all $j$ and for all $t > 0$.

e)  $  (\log A - \log A_j )K = K( \log B - \log B_j)$ for all $j$.

 \end{thm}

 \pf  Clearly (a)  $\imp $ (b).  The implications (b) $\imp $ (c) $\imp $ (d), as well as
 (b)  $\imp $ (a), follow
 from the discussion above.  Differentiation of  (d) at $t = 0$ gives (d) $\imp $ (e), and
  it is straightforward to verify that (e) $ \imp$ (b) with $p = 1$.
Moreover, (d) implies $ \sum_j\tr K^* A_j^{it}KB_j^{1-it}= \tr K^*A^{it}K B^{1-it}$ for all $t$, which implies  (a)  by analytic continuation.  
\qed

 \subsection{Sufficient subalgebras}

When $K=I$, we can obtain a more useful reformulation of the
equality conditions by using results about  sufficient subalgebras
obtained in \cite{JP1,JP2,Pz0}.
Since  the definition and convexity properties of
$J_p(I,A,B)$ extend by continuity to positive semidefinite matrices,
with $\ker B\subseteq \ker A$, we will formulate the conditions
in this more general situation, using the conventions in Section~\ref{sect:not}.
  
Let $N\subseteq M_d$ be a subalgebra, then there is a trace preserving
conditional expectation $E_N$ from $M_d$ onto $N$, such that
$\tr AX=\tr E_N(A)X$ for all $X\in N$. In particular, if 
$N=M_{d_1}\otimes I\subseteq M_{d_1}\otimes M_{d_2}$, then we have
$E_N(A_{12})=\trp_2 A\otimes \tfrac{1}{d_2}I$.

Let $Q_1,\dots,Q_m\in M_d^+$ and assume that $\ker Q_m\subseteq \ker Q_j$ for all $j$.
The subalgebra $N$ is said to be
sufficient for $\{Q_1,\dots, Q_m\}$ if there is a completely positive trace preserving map
$T: N\to M_d$, such that $T(E_N(Q_j))=Q_j$ for all $j=1,\dots,m$. This
definition is due to Petz \cite{Pz0, petz1988} and it is a quantum generalization of the
well known notion of sufficiency from  classical statistics. In \cite{Pz0}, it was
 shown that sufficient subalgebras can be characterized by the condition
 \bee
H(Q_j,Q_m)=H(E_N(Q_j),E_N(Q_m)),\quad \mbox{for all } j
 \eee
 We combine this with the results of the previous section to obtain
  other useful characterizations of sufficiency.

\begin{thm}\label{thm:suff} 
Let $Q_1,\dots, Q_m \in M_d^+$ be such that
$\ker Q_m \subseteq \ker Q_j$ for all $j$.
Let $N\subseteq M_d$ be a subalgebra. The following are equivalent.

(i) $N$ is sufficient for $\{Q_1,\dots, Q_m\}$.

(ii) $E_N(Q_j)^{it}E_N(Q_m)^{-it} P_{(\ker Q_m)^\perp} =Q_j^{it}Q_m^{-it}$, 
for all $j$, $t\in \bf R$.

(iii) There exist $Q_{j,0}\in N^+$,  and $D\in M_d^+$, such that
$\ker D=\ker Q_m$,  and
$ Q_j=Q_{j,0}D $ for $j=1,\dots,m $.

(iv) $J_p(I,Q_j,Q_m)=J_p(I, E_N(Q_j),E_N(Q_m))$ for all $j$ and some $p\in(0,1)$

\end{thm}
The proof of the conditions (i) -- (iii) can be found in \cite{JP1}, see also \cite{MP}. The condition (iv) 
 was proved in \cite{JP2}.

\subsection{Equality conditions with $K=I$}

\begin{thm}\label{thm:eqJpI}
Let $A_1,\dots, A_m$ and $B_1,\dots, B_m$ be positive semi-definite matrices 
with $\ker B_j \subseteq \ker A_j$, and let $A = \sum_j A_j,  B = \sum_j B_j$.
 Then the following are equivalent.

a) $J_p(I,A,B)=\sum_jJ_p(I,A_j,B_j)$ for all $p\in (0,2)$.

b) $J_p(I,A,B)=\sum_jJ_p(I,A_j,B_j)$ for some $p\in (0,2)$.

c) $A_j^{it}B_j^{-it}=A^{it}B^{-it}P_{(\ker B_j)^\perp}$ for all $j$ and $t\in \bf R$

d) There are positive matrices $D_1,\dots, D_m$, with
$\ker D_j=\ker B_j$,  such that
$[A_j,D_j]=[B_j,D_j]=0$, and with $D=\sum_jD_j$ 
\be\label{eq:factoriz2}
A_j=AD^{-1}D_j,\qquad B_j=BD^{-1}D_j
\ee

\end{thm}

\pf As in Section~\ref{sect:Jpeqal}, (b) implies \eqref{basiceq} on $(\ker B_j)^\perp$, with $X_j=B_j, X=B$. This gives
\be
(\Delta_{A_jB_j}+tI)^{-1}(I)=(\Delta_{AB}+tI)^{-1}(I) \qquad \hbox{on} \quad (\ker B_j)^\perp.
\ee
Then (c) follows from the Cauchy integral formula as in Section 
\ref{sect:Jpeqal}.

To  show (c) implies (d), we will use Theorem~\ref{thm:suff}.  First let
  $N=I\ot M_d \subseteq  M_m\ot M_d$ and
let $\wtd  A_{12}$, $\wtd B_{12}$ be the block-diagonal matrices in
$ M_m\ot M_d$, defined by \eqref{ABdiag}.
Clearly, we  have
$\ker  \wtd A_{12}\supseteq \ker  \wtd B_{12}= 
\sum_j \proj{e_j}\ot \ker B_j$ and $E_N(\wtd A_{12})=\frac{1}{m}I\otimes A$,
 $E_N(\wtd B_{12})=\frac{1}{m}I\otimes B$.
Then (c) implies
$E_N(\wtd A_{12})^{it}E_N(\wtd B_{12})^{-it}P_{(\ker \wtd B_{12})^\perp}= 
\wtd A_{12}^{it}\wtd B_{12}^{-it}$ for all $t$.   Then by using Theorem 
\ref{thm:suff} with $Q_1 = \wtd A_{12}, Q_m = Q_2 = \wtd B_{12}$,
we can conclude that there are positive
matrices $A_0,B_0\in M_d$ and $ D_{12}\in (M_m\ot M_d)^+$, such that
$\ker  D_{12}=\ker \wtd B_{12}$, $[I\ot A_0, D_{12}]=[I\ot B_0, D_{12}]=0$
and
\be  \label{ctod1}
\wtd A_{12}=(I\ot A_0) D_{12},\qquad  \wtd B_{12}=(I\ot B_0) D_{12}
\ee 
Since $\wtd  A_{12}$, $\wtd  B_{12}$ are block diagonal,  
$ D_{12} =\sum_j  \proj{e_j}\ot D_j$  must also be block diagonal
with $D_j\in M_d^+$,
$\ker D_j=\ker B_j$, $[A_0,D_j]=[B_0,D_j]=0$ for all $j$ and
\be   \label{ctod2}
A_j=A_0D_j,\qquad B_j=B_0D_j.
\ee
Taking $\trp_1$ in \eqref{ctod1} gives $A = A_0 D$ and $ B = B_0 D$. 
Using this in \eqref{ctod2} gives \eqref{eq:factoriz2} which proves (d).
 The implications (d) $\imp$ (a) $\imp$ (b) are straightforward.
\qed

We return briefly to the case of arbitrary $K$.
Note that if the condition (d) holds and $[D_j,K]=0$ for all $j$, then  
$ J_p(K,A,B) = \sum_j  J_p(K,A_j,B_j)$  for all $p \in (0,2)$, 
this gives a sufficient, but not necessary, condition for equality if $K\ne I$.
The next result reduces the case of $K$ unitary to $K = I$.  Then, we
can apply the conditions of Theorem~ \ref{thm:eqJpI} to $A_j$
and $K B_j K^*$.
\begin{thm}  \label{thm:unit}
If $K$ is unitary, then $ J_p(K,A,B) =  \sum_j  J_p(K,A_j,B_j)$ if and only if
$ J_p(I,A,KBK^*) = \sum_j  J_p(I,A_j,KB_jK^*) $
\end{thm}
 \pf  When $K$ is  unitary, then  $K B^p K^* =  (K B  K^*)^p $ which implies
$J_p(K,A,B) =  J_p(I,A,KBK^*) $.
  \qed

One can try to extend the results of this section to the case $\|K\|\le 1$, and
 hence to all $K$, by using the unitary dilation 
 \bee
\mathcal U = \pmx K & L\\ -L & K \emx
\eee
where $L=U(1-|K|^2)^{1/2}$ and $K=U|K|$ is the polar decomposition. Then, with
 \bee
\ca = \pmx A & 0\\ 0 & 0 \emx,\quad \cb = \pmx B & 0 \\ 0 & 0 \emx
\eee
we have $J_p(K,A,B)=J_p(\mathcal U, \ca,\cb)$, so that we may use 
Theorem~\ref{thm:unit} to get conditions for equality.  But note that the conditions of Theorem~\ref{thm:eqJpI} require that $\ker \mathcal U\cb_j\mathcal U^*
\subseteq \ker \ca_j$ and it can be shown that this implies 
$P_{(\ker A_j)^\perp} K P_{(\ker B_j)^{\perp} }K^* =  P_{(\ker A_j)^\perp}$,
where $P_{\cal N}$ denotes a projection onto the subscripted space. In particular,
 if all $A_j$ and $B_j$ are invertible, this restricts us to unitary $K$.

\subsection{Equality in monotonicity under partial trace}
 
 It is easy to see that when $A_{12} = A_1 \ot A_2$ and $B_{12} = B_1 \ot B_2$,  then
 $J_p(I,A_{12},B_{12}) = J_p(I,A_2,B_2)$ if and only if $A_1 = B_1$
 with $\tr A_1 = 1$.   However, it is not necessary that  
 $A_{12} = A_1 \ot A_2$.
 The equality conditions are given by the following theorem.
  \begin{thm}\label{thm:eqpartial} 
  Let $K_{12}=I_{12} $ and $A_{12},B_{12}\in \cb(\ch_1\ot\ch_2)^+$, with $\ker B_{12}\subseteq \ker A_{12}$.
 Equality holds in \eqref{Jpmono} if and only if

   (i)   $\ch_2 = \bigoplus_n \ch_n^L  \ot \ch_n^R $,

   (ii)  $A_{12} =  \bigoplus_n   A_n^L \ot A_n^R $ with 
   $A_n^L\in\cb(\ch_1\ot \ch_n^L)^+$ and $A_n^R\in \cb(\ch_n^R)^+ $,

   (iii)  $B_{12} =  \bigoplus_n   B_n^L \ot B_n^R $ with $B_n^L\in\cb(\ch_1\ot 
   \ch_n^L)^+$ and $B_n^R\in \cb(\ch_n^R)^+  $,

        (iv)  $ A_n^L  =  B_n^L $ for all $n$.
    \end{thm}
 \pf
 Let us denote $A_j=\tfrac{1}{d_1}\cw_jA_{12}\mathcal \cw_j^*$,
$B_j=\tfrac{1}{d_1}\cw_jB_{12}\cw_j^*$, with $\cw_j$ defined as in
the proof of Theorem \ref{thmJpmono}. Then we  get that equality in
\eqref{Jpmono} is equivalent to
$$J_p(I_{12},\sum_jA_j,\sum_j B_j)=\sum_jJ_p(I_{12},A_j,B_j)$$
 By  Theorem~\ref{thm:eqJpI}, equality for
 some $p$ implies equality for all $p$, so that $J_p(I_{12},A_{12},B_{12})=
 J_p(I_2,\trp_1\, A,
 \trp_1\, B)=J_p(I_{12},E_N(A_{12}),E_N(B_{12}))$ for $p\in (0,1)$, where $N$ is the subalgebra 
 $I_1\ot \cb (\ch_2)\subseteq \cb (\ch_1\ot \ch_2)$. Hence $N$ is sufficient for 
 $\{A_{12},B_{12}\}$ and, by Theorem \ref{thm:suff},
there are some $A_R,B_R\in 
 \cb (\ch_2)^+$ and $D\in \cb (\ch_1\ot \ch_2)^+$, $\ker D=\ker B_{12}$, such that
$ [(I_1\ot A_R),D]=[(I_1\ot B_R),D]=0$ and 
\be\label{eqpart}
 A_{12}=D(I_1\ot A_R),\quad  B_{12}=D(I_1\ot B_R)
\ee
Now let $M_1$ be the subalgebra in $\cb (\ch_2)$, generated by
$A_R,B_R$. Then $D\in (I_1\ot M_1)^\prime =\cb (\ch_1)\otimes M_1^\prime$
where  $M^\prime$ denotes the commutant of $M$.  There
is a decomposition $\ch_2=\bigoplus_n \ch^L_n\otimes
\ch_n^R$, such that
$$M_1'=\bigoplus_n \cb (\ch^L_n)\otimes 1_n^R, \qquad
M_1=\bigoplus_n 1^L_n\otimes \cb(\ch^R_n)$$ and $D=\bigoplus_n
D_n\ot 1_n^R$, where $D_n\in \cb (\ch_1\ot \ch_n^L)$. Since 
$A_R,B_R\in M_1$, we get the result, with
$A_n^L=B_n^L=D_n$.
The converse  can be verified directly.
\qed

    Applying this result in the   case $A_{12} \mapsto A_{123}$ and
    $B_{12} \mapsto   A_{12} \ot I_3$ gives equality conditions
in \eqref{Jpssa}.   Since these are independent of $p$, they are 
identical to the conditions,  first given in \cite{HJPW}, for equality in  SSA \eqref{ssa} 
which corresponds to $p = 1$.   
    \begin{cor}  \label{cor:eqssa}
    Equality holds in \eqref{Jpssa} if and only if

     (i)   $\ch_2 = \bigoplus_n \ch_n^L  \ot \ch_n^R $.

    (ii)  $A_{123} =  \bigoplus_n   A_n^L \ot A_n^R $ with $A_n^L \in \cb(\ch_1 \ot \ch_n^L)  $
   and $ A_n^R \in  \cb(\ch_n^R \ot \ch_3)$
    \end{cor}

\pf
It suffices to let $A_{12}\to A_{123}$ and $B_{12}\to A_{12}\otimes I_3$
 in Theorem \ref{thm:eqpartial}.
\qed

To apply these results in Section~\ref{sect:LC},
it is  useful to observe that  condition (ii) in   Corollary~\ref{cor:eqssa} above 
can be written as 
\be\label{eqssa2}
A_{123} =(F_L\otimes I_3)(I_1\otimes F_R)
\ee
with
 $F_L\in \cb (\ch_1\ot \ch_2)^+$, $ F_R\in \cb (\ch_2\ot\ch_3)^+$,
$[F_L\otimes I_3, I_1\otimes F_R]=0$.   Combining this with part (d) of Theorem~\ref{thm:eqJpI}
gives the following useful result, which essentially allows us to bypass
the need to apply Theorem~\ref{thm:eqJpI}  to $J_p(I, A_j, \cw_n A_j \cw_n)$.
\begin{cor}\label{cor:ssas} Let $A_j\in M_{d_1}\otimes M_{d_2}$, $A=\sum A_j$. Then
\begin{equation}\label{cor2}
J_p(I_{12},A, (\trp_2 A)\otimes I_2)=\sum_j J_p(I_{12}, A_j, (\trp_2 A_j)\otimes I_2)
\end{equation}
if and only if there are $D_j\in M_{d_1}^+$,  such that
$\ker D_j=\ker \trp_2 A_j$, $[A_j, D_j\otimes I]=0$ and
$A_j=A(D^{-1}D_j\otimes I)$ with $D=\sum_j D_j$.

\end{cor}

\pf Let $ \wtd A_{123}=\sum_j \proj{e_j}\otimes A_j \in  M_m \otimes M_{d_1}\otimes M_{d_2}$, 
 then  $A = \wtd A_{23} \in M_{d_1}\otimes M_{d_2}$ and \eqref{cor2} can be written as
$$
J_p(I_{23}, \wtd A_{23}, \wtd A_2\otimes I_3) = J_p(I_{123}, \wtd A_{123}, \wtd A_{12}\otimes I_3)
$$
By  \eqref{eqssa2}, this is equivalent to the existence of
$F_L$ and $F_R$, $[(F_L\ot I_3),(I_1\ot F_R)]=0$, such that $\wtd A_{123} =
(F_L\otimes I_3)(I_1\otimes F_R)$.
Since $\wtd A_{(1)(23)}$ is block-diagonal, $F_L$ must be of the form
$F_L=\sum_j \proj{e_j}\otimes D_j$, so that $A_j=F_R (D_j\otimes I)$.
Then $\trp_2 A_j=D_j\trp_2 F_R$ which implies that $\ker D_j \subseteq \ker \trp_2 A_j$.
If we let $P_j= P_{(\ker \trp_2 A_j)^\perp}$, then $P_j$ commutes with $D_j$ and 
\[
A_j= (P_j\otimes I)A_j=(P_jD_j\otimes I)F_R,
\]
so that we can assume that $\ker D_j=\ker \trp_2 A_j$, by taking $P_jD_j$ instead of $D_j$.
Taking $\trp_1$ of \eqref{eqssa2} gives $A = (D \ot I_3) F_R = F_R (D \ot I_3) $ 
 so that $A_j = A (D^{-1} D_j \ot I)$.
\qed

  \section{Equality in joint convexity of Carlen-Lieb}  \label{sect:LC}

 Carlen and Lieb \cite{CL2} obtained several convexity inequalities from
 those  of the map
 \be
     \Upsilon_{p,q}(K,A) \equiv \tr (K^* A^p K)^{q/p}
 \ee
 using an identity which we write only for $q = 1$ and $p > 1$
 in our notation as
 \be   \label{upsopt}
    \Upsilon_{p,1}(K,A) =  (p-1) \inf \big\{ J_p(K,A, X) + \tfrac{1}{p}\tr X+
    \tfrac{1}{p(p-1)} \tr K^* A K  : X > 0  \big\}
 \ee
We introduce the closely related quantity
\be
   \wh{ \Upsilon}_{p,1}(K,A)  & = &   \inf \big\{ J_p(K,A, X) +  \tfrac{1}{p} \tr  X  : X > 0  \big\} \\
                 & = &   \tfrac{1}{(p-1)}  \big(  \Upsilon_{p,1}(K,A)  - \tfrac{1}{p}\tr K^* A K \big)
\ee
which is well-defined for all $p \in (0,2)$ and allows us to continue to treat the
cases $p < 1$ and $p > 1$ simultaneously, as well as include the special
case $p = 1$ for which
\be
      \wh{ \Upsilon}_{1,1}(K,A) & = & - \tr K^* A K \log (K^* A K) + \tr K^* (A \log A ) K  + \tr K^* A K  \nn \\
                           & = & S(K^* A K) + \tr K K^* A \log A   + \tr K^* A K 
\ee

 Since we are dealing with finite dimensional spaces, the infimum
in \eqref{upsopt} has a minimizer which satisfies
 \be   \label{xmin}
       X_{\min} = (K^* A^p K)^{1/p}.
 \ee
 For fixed $K$,  let $X_j$ denote the minimizer
associated with $A_j$.   Then
 \be
     \wh{\Upsilon}_{p,1}(K,A_1) + \wh{\Upsilon}_{p,1}(K,A_2)   & = &
          J_p(K,A_1,X_1) + \tfrac{1}{p}\tr X_1 +  J_p(K,A_2,X_2)  +\tfrac{1}{p}
	  \tr X_2  \qquad  \nn \\
      &     \geq  & J_p(K, A_1 + A_2, X_1+X_2) + \tfrac{1}{p}\tr (X_1 + X_2)    \\   \nn
         &     \geq  &  \inf \big\{ J_p(K,A_1 + A_2,X) + \tfrac{1}{p}\tr X  : X > 0  \big\} \\
         & =  &       \wh{\Upsilon}_{p,1}(K,A_1 + A_2)
\ee
which proves convexity of $\wh{\Upsilon}_{p,1}$. Note that equality above requires both 
 $X=\sum_j X_j$ and $J_p(K,A,X)=\sum_jJ_p(K,A_j,X_j)$,
where $X$ is the minimizer associated with $A$.



Now  we introduce some notation following the strategy
in the published version of  \cite{CL2}.
Let $|  \one \ket$ denote the vector  $(1 , 1, \ldots , 1)$ with all components
$1$ and  $|e_1  \ket$ the vector $(1 , 0, \ldots , 0)$.   Define
\be
     \ck =  \tfrac{1}{d}  I \ot  | \one \kb e_1 | = \pmx I & 0 & \ldots & 0 \\ I & 0 & \ldots  & 0 \\ \vdots & \vdots & & \vdots \\
        I & 0 & \ldots & 0 \emx
        \ee
 and
  \be
        \ca_j =  \sum_k  A_{jk}  \ot \proj{e_k}  =  \bigoplus_k A_{jk} =
        \pmx A_{j1} & 0 & 0 & \ldots & 0 \\ 0& A_{j2} & 0 & \ldots  & 0 \\
            0 & 0 & A_{j3} & \ldots & 0 \\  \vdots & &  & \ddots & \vdots\\   \emx,
\ee
 and $\ca = \sum_j \ca_j =  \sum_k A_k\ot \proj{e_k} =    \oplus_k A_k$ with $  A_k = \sum_j A_{jk}$. Then
\bee
\ck^*\ca^p\ck= \big(\sum_k A_k^p \big) \ot \proj{e_1} 
\eee
 With this notation, we make some
definitions following Carlen and Lieb but modified
 to  allow a unified treatment of $p \in (0,2)$.
\be
 \Phi_{(p,1)}(\ca) =  \Phi_{(p,1)}(A_1,  A_2, A_3 \ldots ) &   \equiv  &  {\Upsilon}_{p,1}(\ck,\ca)    \\
     &  =  &   \tr \big( A_1^p + A_2^p + A_3^p + \ldots \big)^{1/p} \nn  \\
   \wh{\Phi}_{(p,1)}(\ca) =    \wh{\Phi}_{(p,1)}(A_1,  A_2, A_3 \ldots )
      &   \equiv  & \wh{\Upsilon}_{p,1}(\ck,\ca)   \\ \nn
     & = &   \tfrac{1}{(p-1)} \Big[  \Phi_{(p,1)}(A_1,  A_2, A_3 \ldots ) - 
     \tfrac{1}{p}\sum_k \tr A_k  \Big]  \qquad
 \ee
 The definitions of $\Phi$ and $\wh{\Phi}$ apply only when $\ca$
  is a block diagonal matrix in $M_{d_1} \ot M_{d_2}$.   We now extend this
  to an arbitrary matrices $\ca_{12} \in M_{d_1} \ot M_{d_2}$.
  \be   \label{psidef}
       \Psi_{(p,1)}(\ca_{12} ) &   \equiv  &   \trp_1 \,   \big( \trp_2 \, \ca_{12}^p \big)^{1/p}   \\
       \wh{\Psi}_{(p,1)}(\ca_{12}) &   \equiv  &
        \tfrac{1}{(p-1)} \big[  \Psi_{(p,1)}(\ca_{12}) -  \tfrac{1}{p}\tr \ca_{12}  \big]     \label{psihat}
\ee
For $p = 1$, the formulas with hats are related to the conditional entropy, from which they differ
by a constant 
\be   \label{phi1}
   \wh{\Phi}_{(1,1)}(A_1,  A_2, A_3 \ldots ) - \tr  \ca_{12} & = &
     - \tr \textstyle{\big(\sum_k A_k  \big)  \log \big(\sum_k A_k  \big)  } + \sum_k A_k \log A_k  \nn \\
       & = &  \textstyle{ S\big(\sum_k A_k  \big) } - S(\ca_{12})  = J_1\big(I, \ca_{12}, \trp_2 \, \ca_{12} \ot I_2  \big) \qquad \\
      \wh{\Psi}_{(1,1)}(\ca_{12})   - \tr  \ca_{12}   & = & S(\ca_1) - S(\ca_{12}) = H(\ca_{12},\ca_1\ot I_2)   \label{psi1}
\ee

 When $\ca_{12}$ is block diagonal,  $ \Psi_{(p,1)}(\ca_{12} ) =\Phi_{(p,1)}(\ca_{12}   ) $ with the
 understanding that $\trp_2 \ca_{12}  = \sum_k A_k$.
Now let $W_n$ denote the generalized Pauli matrices as in Section~\ref{sect:mono}, $\cw_n = I_1 \ot W_n$ and define
\be  \label{A123blk}
     \ca_{123} = \sum_n  \cw_n \ca_{12}    \cw_n^*  \ot \proj{e_n} = \bigoplus_n   \cw_n \ca_{12}    \cw_n^*
\ee
so that $  \ca_{123}$ is block diagonal with blocks $\cw_n \ca_{12}   \cw_n^*$.
Then
\be \label{psiblock}
d_2^{\frac{1+p}{p}} \Psi_{(p,1)}(\ca_{12})=  \Phi(\ca_{(12)(3)})
    =   \Phi\big(\cw_1 \ca_{12}   \cw_1^* , \,
   \cw_2 \ca_{12}    \cw_2^*, \, \ldots \big).
   \ee
 It is straightforward to show that for $p \in (0,2)$ the functions
$ \wh{ \Phi}_{(p,1)}(\ca) $ and $ \wh{ \Psi}_{(p,1)}(\ca)$
are all convex in $\ca$,  inheriting this property from the quantities from which
they are defined.
In view of \eqref{phi1} and \eqref{psi1}, the conditions for equality in the
next two theorems are not surprising.
\begin{thm}   \label{thm:phieq}
The function   $ \wh{\Phi}_{(p,1)}(\ca) $ is
convex in $\ca$ for $p \in (0,2)$.   Moreover, the following are equivalent:

(i)   $ J_p\big(I, \ca, (\trp_2 \, \ca) \ot I_2  \big) = \sum_j J_p \big(I, \ca_j, ( \trp_2 \, \ca_j )  \ot I_2 \big) $,

(ii)  There  are matrices $D_j > 0$, $D=\sum_jD_j$, such that $[A_{jk},D_j] =0$,
 $\ker D_j =\ker (\sum_k A_{jk})$ and   $A_{jk}=A_kD^{-1}D_j$,

(iii)  $\wh  \Phi_{(p,1)}(A_1,  A_2, A_3 \ldots )  = \sum_j  \wh \Phi_{(p,1)}(A_{j1},  A_{j2}, A_{j3} \ldots )  $.

\end{thm}

\pf It follows from Corollary~\ref{cor:ssas} and the fact that $\ca_j$ are
 block-diagonal that (i) $\iff$ (ii) 
and it is straightforward to verify
that (ii) $ \imp  $ (iii). Moreover, (iii) implies (i) for $p=1$, by
\eqref{phi1}.  To show that (iii) implies (ii) for $p\ne 1$,
observe that  (iii), implies  $\wh \Upsilon_{p,1}(\ck,\ca)=
\sum_j \wh \Upsilon_{p,1}(\ck,\ca_j)$,  and this implies
\be\label{eq:eqt}
J_p(\ck,\ca,\mathcal X)=\sum_j J_p(\ck,\ca_j,\mathcal X_j)
\ee
where  
$\mathcal X_j=(\ck^* \ca^p_j\ck)^{1/p}=
X_j\otimes \proj{e_1}$ and 
$\sum_j \mathcal X_j = \mathcal X=(\ck^*\ca^p\ck)^{1/p}= \linebreak X\otimes \proj{e_1}$, 
with $X_j=(\sum_k A_{jk}^p)^{1/p}$ and $X=(\sum_k A_k^p)^{1/p}$.
Since
\bee
 \ck^* \ca_j^p\ck\mathcal X_j^{1-p}=\sum_k A_{jk}^pX_j^{1-p}\otimes \proj{e_1},
\eee
with a similar expression for $ \ck^* \ca^p\ck\mathcal X^{1-p}$, we find
\bee
\sum_k J_p(I,A_k,X)=J_p(\ck, \ca,\mathcal X)=
\sum_jJ_p(\ck, \ca_j,\mathcal X_j)=\sum_{k,j} J_p(I,A_{jk},X_j)
\eee
Convexity then implies that we must have
\be  J_p(I,A_k,X)=\sum_j J_p(I,A_{jk},X_j)  \qquad   \forall ~ k.
\ee
Since $\ker X_j\subseteq \ker A_{jk}$,  Theorem~\ref{thm:eqJpI}  implies that
\begin{equation}\label{eq:eq}
A_k^{it}X^{-it}P_{(\ker X_j)^\perp}= A_{jk}^{it}X_j^{-it}, \quad \mbox{ for all } k,j, t
\end{equation}
After writing $\wtd A_k=\sum_j\proj{e_j}\ot A_{jk}$, $\wtd X=\sum_j \proj{e_j}\ot X_j$, this reads
\bee
\wtd A_k^{it}\wtd X^{-it}=(\tfrac{1}{m}I\ot \trp_1 \wtd A_k)^{it}(\tfrac{1}{m}I\ot \trp_1 \wtd X)^{-it}P_{(\ker \wtd X)^\perp},
\eee
so that, by Theorem \ref{thm:suff}, there are   elements $B_k\in M_d^+$
 and $D\in (M_m\ot M_d)^+$, such that  $\ker D = \ker \wtd{X} $, $[(I\ot B_k),D]=0$ and 
 $\wtd A_k=(I\ot B_k)D$.   As before, one finds
 $D=\sum_j \proj{e_j}\ot D_j$ for some   $D_j\in M_d^+$ which implies (ii).
 \qed

\begin{thm}   \label{thm:psieq}
The function   $ \wh{\Psi}_{(p,1)}(\ca_{12}) $ is
convex in $\ca_{12}$ for $p \in (0,2)$.   Moreover, if we let 
 $  \ca_{123}$ denote the  block diagonal matrix with blocks 
$ \cw_n  \ca  \cw_n^* $, the following are equivalent:

 (i) $ J_p( I, \ca_{123}, \ca_1 \ot I_{23}) = \sum_j J_p \big( I, (\ca_{123})_j , (\ca_1)_j   \ot I_{23}\big) $
with $\ca_{123}$ defined by \eqref{A123blk},
 
(ii)  There are matrices $D_j\in M_{d_1}^+$, $D=\sum_j D_j$, such that  $\ker D_j = \ker( \ca_1)_j$, 
 $[\ca_j, D_j\ot I] =0$ and 
$\ca_j=\ca(D^{-1}D_j\ot I)$.

(iii) $\wh  \Psi_{(p,1)}(\ca) = \sum_j  \wh  \Psi_{(p,1)}(\ca_j) $.
\end{thm}
\pf  It follows from the definition of    $  \ca_{123}$, that
$d_2^{\frac{1+p}{p}} \Psi_{(p,1)}(\ca) = \Phi(\ca_{123})$. The equivalence (i) $\iff$ (iii)
follows immediately from Theorem~\ref{thm:phieq}, and  (i) $\iff$ (ii) can be shown to follow  from
  Corollary \ref{cor:ssas}. \qed

\begin{thm}  \label{thm:psimono}
The following monotonicity inequalities hold,
\bsq   \label{psimono} \begin{align}
    &\wh{\Psi}_{(p,1)}(\ca_{23})  \leq     \wh{\Psi}_{(p,1)}(\ca_{123}),\quad      p \in (0,2  ) \\
    &{\Psi}_{(p,1)}(\ca_{23})   \geq   {\Psi}_{(p,1)}(\ca_{123}),\quad      p \in (0,1)    \\
    &{\Psi}_{(p,1)}(\ca_{23})   \leq      {\Psi}_{(p,1)}(\ca_{123}),\quad       p \in [1,2)
 \end{align} \esq
  Moreover, equality holds if and only if the conditions of Corollary~\ref{cor:eqssa}
  are satisfied.
\end{thm}

\pf It suffices to give the proof for $\wh{\Psi}$ since the other
inequalities  follow immediately.   The argument is similar to that for Theorem~\ref{thmJpmono}.
Let $W_n$ denote the generalized Pauli matrices of Section~\ref{sect:mono},
 but now let $\cw_n = W_n \ot I_{23}$.    Then the convexity of  $\wh{\Psi}_{(p,1)}(\ca_{23})$
 implies
 \bee
  \wh{\Psi}_{(p,1)}(\ca_{23})  & = &  \tfrac{1}{d_1}  \wh{\Psi}_{(p,1)}( I_1 \ot \ca_{23}) \\
   & = &   \tfrac{1}{d_1 } \wh{\Psi}_{(p,1)}
        \big(  \tfrac{1}{d_1 }  \textstyle{   \sum_n  \cw_n  \ca_{123}  \cw_n \big) } \\
         & \leq &  \tfrac{1}{d_1^2 }  \sum_n  \wh{\Psi}_{(p,1)}( \cw_n  \ca_{123}  \cw_n )
         ~ =  ~  \wh{\Psi}_{(p,1)}( \ca_{123} )
 \eee
where we used the invariance of $\wh{\Psi}$ under unitaries of the form $U_1 \ot I_{23}$. In the case $p = 1$, it follows from \eqref{psi1} that
$\wh{\Psi}_{(1,1)}(\ca_{23})  \leq \wh{\Psi}{(1,1)}(\ca_{123})$ becomes
\be
   S(\ca_{2}) - S(\ca_{23}) \leq   S(\ca_{12}) - S(\ca_{123})
\ee
which is SSA.  Because the equality conditions in 
Theorem~\ref{thm:psieq}
are independent of $p$, they are identical to those for SSA, which are given in
Corollary~\ref{cor:eqssa}.\qed

The Carlen-Lieb triple Minkowski inequality   for the case $q = 1$ is an
immediate corollary of Theorem~\ref{thm:psimono}.  Observe that
\bsq  \label{phitrip} \be
  \trp_3 \,  \trp_1  \big(  \trp_2 \ca_{123}^p \big)^{1/p} & =  &
              \Psi_{(p,1)}( \ca_{(13),(2)})      \\
    \trp_3 \, \big[ \trp_2 (\trp_1 \,   \ca_{123})^p \big]^{1/p}  & = & \Psi_{(p,1)}( \ca_{32})
         \ee \esq
so that it  follows immediately from (\ref{psimono}c) that
      \be
      \trp_3 \, \big[ \trp_2 (\trp_1 \,   \ca_{123})^p \big]^{1/p} =    \Psi_{(p,1)}( \ca_{32})
           \leq  \Psi_{(p,1)}( \ca_{132})
         =    \trp_3 \,  \trp_1  \big(  \trp_2 \ca_{123}^p \big)^{1/p}
             \ee
for $1 < p \leq 2$ and from  (\ref{psimono}b) that the inequality reverses for $0 < p < 1$.   Moreover,   the conditions for equality are again independent of $p$
 and identical to those for equality in SSA, given in Corollary~\ref{cor:eqssa}.
 
 \section{Final remarks}
 
 It should be clear that the results in Section~\ref{sect:WYDgen} are not restricted to $J_p(K,A,B)$.
 The function $g_p(x)$ given in \eqref{quasi} can be replaced by any
 operator convex function of the form $g(x) = x f(x)$ with $f$ operator monotone on 
 $(0,\infty)$.   Moreover, if the measure $\nu(t) $ in \eqref{intrep} is supported on  $(0,\infty)$,
 then the conditions for equality are identical to those in Section~\ref{sect:eq}.
 
 In particular, our results go through with $g_p$ replaced by $\wtd g_p$  and
 $J_p(I, A,B)$ replaced by  $\wtd J_p(I, A,B)$, which is
 well-defined for $p \in [-1,1)$ with $\wtd J_0(I, A,B) = H(B,A)$.   
 Thus our results can be extended to all $p \in (-1,2)$.    
 The case $p = 2$ 
reduces to the convexity of $(A,X) \mapsto \tr X^* A^{-1} X$ with $A > 0$
 proved in \cite{LbR2}.
One can show that equality holds if and only if $X_j = A_j T \quad  \forall ~ j$
with $T = A^{-1} X$.  We recently learned that  Kiefer  \cite{Kf} proved the $p = 2$ 
convexity, by a different method,  much earlier and also
 found  these equality conditions.
  

 There have been various attempts, e.g., the Renyi \cite{Renyi} and Tsallis \cite{T} entropies, to 
 generalize quantum entropy in a way that gives the usual von Neumann entropy at $p = 1$.
In this paper we have considered two extensions of the conditional entropy involving
an exponent $p \in (0,2)$,
  namely, 
  \begin{itemize}
  
  \item  $J_p(I, A_{12}, A_1)$ which gives 
  $\tr A_{23}^p A_{2}^{1-p}  ~  \begin{array}{c} \leq  \\   \geq  \end{array} ~
     \tr A_{123}^p A_{12}^{1-p}  ~~  \begin{array}{c} p \in (0,1) \\   p \in  (1,2)  \end{array}   $
     and
     can be thought of as a   pseudo-metric; and
  
  \item $\wh{\Psi}_{(p,1)}( \ca_{12}) $ which gives 
  $ \trp_2 ( \trp_3 \ca_{23}^p)^{1/p}  ~  \begin{array}{c} \geq  \\   \leq  \end{array} ~
     \trp_{12} ( \trp_3 \ca_{123}^p)^{1/p}   ~~
      \begin{array}{c} p \in (0,1) \\ p  \in (1,2)  \end{array}  $ 
      and can be thought of as  a pseudo-norm.
  
  \end{itemize}
  These expressions 
 are quite different for $ p \neq 1$, but arise from quantities with the
 same convexity and monotonicity properties, as well as
 the same equality conditions which are independent of $p$.
Moreover,  both yield SSA at $p = 1$ and the
equality conditions for $p \neq 1$ are identical to those for SSA.  This independence
of non-trivial equality conditions  on the precise form of the function seems  remarkable.

If one uses $\wtd g_p   $ and $\wtd J_p(I,A,B)$ from \eqref{wtdg}, then the inequalities
above hold with $ p \in (1,2) $ replaced by $p \in (-1,0)$ and  SSA corresponds to $p = 0$.


  \bigskip
 
\appendix

\section{Proof of the key Schwarz inequality}

For completeness, we include the proof of the joint convexity of
$(A,B,X) \mapsto \linebreak 
\tr X^*  (L_A +  t R_B)^{-1}(X) $ when $A,B >0$ and $t > 0$.   Since this function is 
homogeneous of degree one, it suffices to prove subadditivity.
Now let   
\be   \label{Mj}
M_j = (L_{A_j} + t R_{B_j})^{-1/2}(X_j) -  (L_{A_j} + t R_{B_j})^{1/2}(\Lambda) .
\ee
Then one can verify that
\be  \label{eq:Schz1}
   0  & \leq & \sum_j \tr M_j^{*} M_j  ~ = ~ \sum_j \bra M_j , M_j \ket  \nonumber  \\   
     & = &    \sum_j \tr X_j^{*}  (L_{A_j}  + t R_{B_j})^{-1}(X_j)  - 
     \tr   \big( \ts{ \sum_j } X_j^{*}  \big) \Lambda    
               \\    & ~  & ~ \qquad \qquad  -  \tr \Lambda^{*} \big( \ts{\sum_j X_j } \big)
        +    \tr \Lambda^{*}    \ts{ \sum_j} \big( L_{A_j}  + t R_{B_j}) \Lambda  .   \nonumber 
         \ee
Next, observe that  for any matrix $W$,
\bee 
\sum_j  \big( L_{A_j} + t R_{B_j})(W) =   \sum_j \big( A_j W + t W B_j \big)  
 =  L_{\sum_j A_j}(W)  +  t R_{\sum_j B_j}(W)  .
                \eee
Therefore, inserting the choice
 $\Lambda =   \big(  L_{\sum_j A_j}   +  t R_{\sum_j B_j} \big)^{-1}  
         \big( \ts{\sum_j X_j } \big) $
in  (\ref{eq:Schz1}) yields
 \be    \label{eq:Schwzt}
  \tr   \big( \ts{ \sum_j } X_j \big)^{*}  \dfrac{1}{  L_{\sum_j A_j}   +  t R_{\sum_j B_j} } 
         \big( \ts{\sum_j X_j } \big)
  \leq   \sum_j \tr X_j^{*}  \dfrac{1}{L_{A_j} + t R_{B_j}}(X_j) .
            \ee
for any $t \geq 0$.      \qed

\bigskip


\end{document}